# How to Avoid Both the Repugnant and Sadistic Conclusions without Dropping Standard Axioms in Population Economics

Norihito Sakamoto[*]

## Abstract

This study investigates possibility and impossibility results of the repugnant and sadistic conclusions in population ethics and economics. The repugnant conclusion says that an enormous population with very low well-being is socially better than any smaller population with sufficiently high well-being. The sadistic conclusion says that adding individuals with negative well-being to a society is socially better than adding individuals with positive well-being to it. Previous studies have often found it challenging to avoid both undesirable conclusions. However, I demonstrate that a class of acceptable social welfare orderings can easily prevent these conclusions while adhering to standard axioms, such as anonymity, strong Pareto, Pigou-Dalton transfer, and extended continuity. Nevertheless, if the avoidance requirements for the repugnant and sadistic conclusions are strengthened, it is possible to encounter new impossibility results. These results reveal essential conflicts between the independence axiom and the avoidance of the weak repugnant conclusion when evaluating well-being profiles with different populations.

[*] Tokyo University of Science, E-mail: n-sakamoto@rs.tus.ac.jp
          This study was written during my stay at LSE; I sincerely thank LSE CPNSS for their hospitality and Alex Voorhoeve for his friendship. I am truly grateful for thoughtful comments and kind suggestions from Jason Alexander, Richard Bradley, Campbell Brown, Kensei Nakamura, Koichi Tadenuma, Marcus Pivato, and Alex Voorhoeve. Especially, Marcus and Alex's comments are greatly helpful in revising this paper. I also offer my special thanks for the grants from Fostering Joint International Research (B) (JSPS KAKENHI Grant Number: 20KK0036) and the Joint Research Center for this research.

# 1. Introduction

Population ethics and social choice theory have faced significant challenges in assessing well-being profiles with different populations since the celebrated study by Parfit (1984). Generally, the total utilitarian social welfare ordering has been known to result in the *repugnant conclusion*. To illustrate, let us consider the following two profiles.

$$u_N = \underbrace{(10, \ldots, 10)}_{10 \text{ individuals}},$$

$$v_M = \underbrace{(1, \ldots, 1)}_{101 \text{ individuals}}.$$

Suppose a well-being level of 1 represents a situation in which a person is barely surviving in poverty, while a well-being level of 10 signifies a decent quality of life. If a social planner evaluates the above two profiles using utilitarianism, then the profile $v_M$ (consisting of 101 individuals with a well-being level of 1) would be deemed socially better than the profile $u_N$ (consisting of 10 individuals with a well-being level of 10). This conclusion, where a large population with very low well-being is socially better than any smaller population with sufficiently high well-being, is known as the repugnant conclusion.[1]

On the other hand, if social evaluation is based on *averagism*, $u_N$ with its average well-being of 10 is considered socially better than $v_M$ with its average well-being of 1 in the above example, thus avoiding the repugnant conclusion. However, averagism is known to give rise to another problem known as the *sadistic conclusion*. Suppose now that the threshold of lives worth living is zero, where individuals with positive well-being are enjoying a life worth living, while those with negative well-being are living under a level of life not worth living. Therefore, adding individuals with positive well-being to society should be socially better than adding individuals with negative well-being. Now consider the profile $s_L$, which adds 100 individuals with a well-being level of 1 to $u_N$, and the profile $t_K$, which adds 1 individual with a well-being level of (-1) to $u_N$. In this case, averagism would judge $t_K$ to be socially better than $s_L$, where the addition of individuals with negative well-being is socially better than the addition of individuals with positive well-being. This result is called the sadistic conclusion.[2]

---

[1] See Zuber et al. (2021) for a comprehensive summary of the repugnant conclusion. Greaves (2017) is an excellent survey on social choice problems in population ethics. Note that, as seen from the above example, the repugnant conclusion does not need the hypothetical existence of almost infinite number of individuals. It seems that the difficulty of population ethics lies in the fact that various types of undesirable conclusions can easily emerge even with a relatively small population size.

Arrhenius (2000) proved that it is impossible to avoid the repugnant and sadistic conclusions while adhering to certain desirable axioms. This impossibility result suggests that any social evaluation rule with variable populations will lead to undesirable conclusions in population ethics. Several studies have established various impossibility theorems, leading the field to conclude that there is no universally acceptable social evaluation rule that avoids both the repugnant and sadistic conclusions.

The aim of this study is to review the previous research on these undesirable conclusions and reexamine the theoretical compatibility of standard axioms in population ethics. The study demonstrates that it is possible for a social welfare ordering to satisfy standard axioms (anonymity, strong Pareto, Pigou-Dalton transfer, and extended continuity) while avoiding the two undesirable conclusions. However, regrettably, this possibility result does not provide an acceptable solution to the social evaluation problem in population ethics. For it shows that if the social planner wishes to further avoid undesirable conclusions, the impossibility theorem would immediately arise. Therefore, it appears that some compromises of desirable axioms are necessary to construct an acceptable

---

[2] Besides the sadistic conclusion, averagism also has the problem that it prefers a profile of one individual with high well-being to a profile of many individuals with slightly lower but sufficiently high well-being.

aggregation rule for social evaluation with variable populations.

The main contributions of this study are as follows. Firstly, it demonstrates the existence of a social welfare ordering that avoids both the repugnant and sadistic conclusions while satisfying the standard axioms. Secondly, it identifies a class of axioms that result in impossibility theorems, where no social welfare ordering can avoid the two undesirable conclusions. Thirdly, it shows that even when the undesirable conclusions are extended to more natural versions, the impossibility theorems still hold, indicating that compromises are necessary for social choice problems with variable populations.

The structure of this study is as follows. Section 2 explains the definitions and axioms used in the study. Section 3 presents the possibility and impossibility theorems regarding the undesirable conclusions. Section 4 summarizes the study's results and discusses any remaining issues.

## 2. Basic Definitions and Axioms

This section provides definitions, notation, and axioms. I define $\mathbb{N}$ as a set of natural numbers, $\mathbb{R}$ as a set of real numbers, $\mathbb{R}_+$ as a set of non-negative real numbers, $\mathbb{R}_{++}$ as a set of positive real numbers, and $\mathbb{R}_{--}$ as a set of negative real numbers. I assume that the well-being of each person can be represented as a real number and be interpersonally comparable. A finite sequence of levels of well-being for each person is referred to as a profile, and the domain of all possible profiles with variable populations is denoted by $U = \bigcup_{n \in \mathbb{N}} \mathbb{R}^n$, where $U_{++}$ is the set of profiles consisting of all positive real numbers, and $U_{--}$ is the set of profiles consisting of all negative real numbers. I use the notation $n * u$ to refer to a profile of $n$ individuals, each with a well-being level of $u$. For simplicity, let $N$ be a set of people with $n$ elements, and $u_N$ be a profile on population $N$. Hence, $M$ is a set consisting of $m$ individuals and a profile $v_M$ is a profile on population $M$, etc.

The problem I aim to address is to consider a class of acceptable social evaluation rules with variable populations. To do this, I introduce a binary relation $\succcurlyeq$ on the set of profiles $U$. For all profiles $u_N$ and $v_M$ in $U$, $u_N$ is considered socially at least as good as $v_M$ if and only if $u_N \succcurlyeq v_M$. I assume that the binary relation $\succcurlyeq$ is complete and transitive. In other words, for any $u_N$ and $v_M$, either $u_N \succcurlyeq v_M$ or $v_M \succcurlyeq u_N$, and for any $u_N$, $v_M$, and $s_L$, $u_N \succcurlyeq v_M$ and $v_M \succcurlyeq s_L$ implies $u_N \succcurlyeq s_L$. I refer to this $\succcurlyeq$ as a *social welfare ordering* (SWO), with $\succ$ and $\sim$ representing its asymmetric and symmetric parts, respectively.

Following Blackorby et al. (2005), I define avoidance of two undesirable conclusions as follows.

***Avoidance of the Sadistic Conclusion.*** $\forall u_N \in U$, $\forall v_M \in U_{++}$, $\forall s_L \in U_{--}$, $(u_N, v_M) \succcurlyeq (u_N, s_L)$.

Arrhenius (2000) coined the term "*the sadistic conclusion*," which says that adding individuals with positive well-being to some population is socially worse than adding individuals with negative well-being. To avoid this conclusion, it is necessary to acknowledge that adding individuals with positive well-being always be socially at least as good as adding those with negative well-being.[3]

***Avoidance of the Repugnant Conclusion.*** $\exists\, u_N \in U_{++},\ \exists\, \varepsilon > 0,\ \forall\, m \in \mathbb{N},\ u_N \succcurlyeq m * \varepsilon.$

Parfit's repugnant conclusion states that for any population consisting of individuals with positive well-being, there exists a population consisting of a huge number of people with a small amount of well-being, which will be socially better than the former population, no matter how sufficiently high the well-being of the population is. To avoid this conclusion, it is required that, for some profile, a population where all individuals have a low level of well-being is socially worse or equal, no matter how large the population size is.[4] Note that the avoidance of these undesirable conclusions can be made weaker, but since it is not a very essential argument, I will not consider such a weaker version of the avoidance here.

---

[3] Some people may feel that this condition is too strong. In fact, it requires that adding a large number of individuals with barely worthwhile positive well-being should be socially better than adding one individual with slightly negative well-being. However, noted that even if this requirement were applied to only comparisons of profiles between the addition of *m* or (*m*+1) individuals with positive well-being and the addition of *m* or (*m*+1) individuals with negative well-being, the main results obtained in this paper would remain the same.

[4] Recently, Zuber et al. (2021) argued that avoiding the repugnant conclusion should not itself be an exclusive and main objective in search for acceptable solutions in population ethics. This paper aims to examine theoretically whether the undesirable conclusions can be avoided, and is silent on whether theory of population ethics should avoid these conclusions.

This paper requires the conditions of anonymity, equity, efficiency, and continuity as standard axioms to be imposed on social welfare orderings. I will also consider weaker versions of these axioms to strengthen impossibility theorems. The definitions of these axioms basically follow Blackorby et al. (2005).

First, as an axiom of procedural impartiality, consider the *anonymity* condition. This requires that the well-being of all individuals should be treated equally.

***Anonymity.*** $\forall$ bijections $\pi$ on $N$, $\forall u_N \in U, u_N \sim u_{\pi(N)}$.

Next, I will introduce two axioms that consider distributional equity. One is the well-known *Pigou-Dalton Transfer*, which requires that transferring the same well-being from the rich to the poor should not decrease social welfare whenever everyone else does not affected.

***Pigou-Dalton Transfer.*** $\forall u_N, v_N \in U, \forall \varepsilon \in \mathbb{R}_{++}$, if $\exists i, j \in N, v_i - \varepsilon = u_i \geq u_j = v_j + \varepsilon$ and $\forall k \in N \setminus \{i,j\}, u_k = v_k$, then $u_N \succcurlyeq v_N$.

The second distributive equity axiom is *Minimal Equity*, which requires that for any given profile, the perfectly equal distribution is at least as good as the original distribution if the sums are the same. This is imposed as the minimal axiom of distributive equity to strengthen impossibility theorems.

***Minimal Equity.*** $\forall u_N \in U$, $n * \bar{u} \succcurlyeq u_N$, where $\bar{u}$ is a mean of $u_N$.

As one category of standard axioms, let us consider a class of efficiency or monotonicity requirements. The first one is *Strong Pareto*, which requires that social welfare should be improved whenever at least someone's well-being is better off and no one else is worse off.

***Strong Pareto.*** $\forall u_N, v_N \in U^N$, if $u_N \geqq v_N$, then $u_N \succcurlyeq v_N$. Moreover, if $u_N > v_N$, then $u_N \succ v_N$.[5]

The second efficiency requirement is *Minimal Increasing*, which requires that in two populations with perfect equality and the same number of individuals, if one population has higher well-being, then that population generates higher social welfare.

***Minimal Increasing.*** $\forall a, b \in \mathbb{R}$ with $a > b, \forall n \in \mathbb{N}, n * a \succ n * b$.

Finally, I introduce the standard continuity axiom. The continuity axiom here requires that both the upper and lower contour sets are closed in a comparison of profiles with different population sizes. That is, a profile $v_M$ on population $M$ is socially better than a profile $u_N$ on population $N$, then so is $v'_M$ that is sufficiently close to $v_M$.[6]

---

[5] For all profiles $u_N, v_N \in U$, [$u_N \geqq v_N$ iff $u_i \geq v_i$ for all $i$] and [$u_N > v_N$ iff $u_i \geq v_i$ for all $i$ and $u_j > v_j$ for some $j$] and [$u_N \gg v_N$ iff $u_i > v_i$ for all $i$].

**Extended Continuity.** For all $n, m \in \mathbb{N}$, for all $u_N \in \mathbb{R}^n$, the sets $\{v_M \in \mathbb{R}^m | v_M \succcurlyeq u_N\}$ and $\{v_M \in \mathbb{R}^m | u_N \succcurlyeq v_M\}$ are closed in $\mathbb{R}^m$.

Finally, consider a non-standard axiom. This axiom requires that social welfare must be unaffected or increased by one addition of zero well-being provided well-being levels of the original population never decrease.[7]

**Monotonicity for the Addition of Zero Well-being**: $\forall u_N \in U_{++}, \exists v_N \in U_{++}$ with $v_N \geqq u_N$ such that $(v_N, 0) \succcurlyeq u_N$.

---

[6] Note that if critical levels exist for all profiles, the continuity requirement extended to profiles with any population size would be the same as extended continuity. Suppose that for any profile $u_N$, there exists some real number $c$ such that a profile $u_N$ is socially indifferent from a profile $(u_N, c)$. Let $c$ be called the critical level of $u_N$. Given the extended continuity, the fact that the set $\{v_M \in \mathbb{R}^m | v_M \succcurlyeq u_N\}$ is closed in $\mathbb{R}^m$ is equivalent to the fact that both the sets $\{v_M \in \mathbb{R}^m | v_M \succcurlyeq (u_N, c_1, \ldots, c_{m-n})\}$ with $m>n$ and $\{v_M \in \mathbb{R}^m | (v_M, c_1, \ldots, c_{n-m}) \succcurlyeq u_N\}$ with $n>m$ are closed in $\mathbb{R}^m$, where $(u_N, c_1, \ldots, c_{m-n}) \sim u_N$ and $(v_M, c_1, \ldots, c_{n-m}) \sim v_M$. The same holds for the lower contour set, so "continuity assuming the existence of critical levels" is equivalent to extended continuity.

[7] Underlying this axiom, at least, is the idea that considers the addition of zero well-being to be neutral with respect to social welfare. Indeed, avoidance of the sadistic conclusion is also based on the similar idea that the addition of individuals with zero well-being should be socially neutral. Some people (especially those who defend the egalitarian view) would never consider such an axiom attractive. In fact, this monotonicity condition requires that when one individual with zero well-being is added to a profile in which each individual enjoys very high well-being, provided that well-being of every individual except the worst-off individual sufficiently increases, the profile is socially better than the original profile, no matter how significant inequalities among them arise.

This monotonicity axiom requires that if an individual with a well-being level of zero is added to a population where all persons have positive well-being, the addition should not be socially worse than the original situation if the well-being level of the original population sufficiently increases.[8]

## 3. Possibility and Impossibility Theorems

In this section, I investigate the theoretical relationships between standard axioms and undesirable conclusions. First, let us clarify the implication of Extended Continuity combined with Avoidance of the Sadistic Conclusion.

Lemma 1. *If an SWO $\succcurlyeq$ satisfies Extended Continuity and Avoidance of the Sadistic Conclusion, then* $\forall u_N \in U, \ \forall m \in \mathbb{N}, \ (u_N, 0) \sim (u_N, m*0)$.

Proof. By Avoidance of the Sadistic Conclusion, it holds that $\forall k, m \in \mathbb{N}, (u_N, 1/k) \succcurlyeq (u_N, m*(-1/k))$ & $(u_N, m*1/k) \succcurlyeq (u_N, -1/k)$.

---

[8] This axiom can be replaced by the following condition to lead to an impossibility result.

**Trade-off Condition.** $\forall u_N \in U_{++}, \ \exists v_{N\text{-}\{i\}}$ with $v_{N\text{-}\{i\}} \geqq u_{N\text{-}\{i\}}$ such that $(v_{N\text{-}\{i\}}, 0) \succcurlyeq u_N$.

The trade-off condition requires that even if one person in a profile with positive well-being is reduced to zero well-being, social welfare is no worse off than before by sufficiently increasing everyone else's level of well-being. Of course, there is an ethical problem with a situation where the worst-off individual makes a great sacrifice for all the rest to gain benefits greatly. Therefore, I will proceed with the discussion by only asking for the monotonicity axiom for the addition of individuals with zero well-being to a population.

By Extended Continuity, $(u_N, 0) \succcurlyeq (u_N, m * 0)$ & $(u_N, m * 0) \succcurlyeq (u_N, 0)$.

Hence, it follows that $(u_N, 0) \sim (u_N, m * 0)$. ∎

The following impossibility result is easily obtained by using the above lemma.

*Theorem 1. There exists no SWO that satisfies Avoidance of the Sadistic Conclusion, Avoidance of the Repugnant Conclusion, Minimal Equity, Minimal Increasing, Extended Continuity, and Monotonicity for the Addition of Zero Well-being.*

Proof. By Monotonicity for the Addition of Zero Well-being, $\forall u_N \in U_{++}$, $\exists v_N \in U_{++}$ with $v_N \geqq u_N$, $(v_N, 0) \succcurlyeq u_N$.

By Lemma 1, $\forall m \in \mathbb{N}$, $(v_N, 0) \sim (v_N, m * 0)$.

Minimal Equity implies that $(m+n) * \frac{1}{m+n}\sum_{i \in N} v_i \succcurlyeq (v_N, m * 0)$.

Since $\lim_{m \to \infty} \frac{1}{m+n}\sum_{i \in N} v_i = 0$ and $\frac{1}{m+n}\sum_{i \in N} v_i$ is strictly decreasing with respect to $m$, for any $\varepsilon > 0$, there exists a natural number $m'$ such that $\varepsilon > \frac{1}{m'+n}\sum_{i \in N} v_i$.

By Minimal Increasing, $(m'+n) * \varepsilon \succ (m'+n) * \frac{1}{m'+n}\sum_{i \in N} v_i$

Thus, $(m'+n) * \varepsilon \succ (m'+n) * \frac{1}{m'+n}\sum_{i \in N} v_i \succcurlyeq (v_N, m' * 0) \sim (v_N, 0) \succcurlyeq u_N$.

Transitivity implies $(m'+n) * \varepsilon \succ u_N$, that contradicts Avoidance of the Repugnant Conclusion.

Note that the proof of this impossibility theorem does not require completeness of SWOs. In fact, the impossibility result still holds in the setting of social *quasi-ordering*. From Theorem 1, if the social planner wants to avoid the two undesirable conclusions without dropping the standard axioms, she needs to drop Monotonicity for the Addition of Zero Well-being. Indeed, dropping Monotonicity for the Addition of Zero Well-being can yield the following possibility result.[9]

*Theorem 2. There exists an SWO that satisfies Avoidance of the Sadistic Conclusion, Avoidance of the Repugnant Conclusion, Anonymity, Pigou-Dalton Transfer, Strong Pareto, and Extended Continuity.*

Proof. Consider the following SWO. For all profiles $u_N, v_M \in U$,

$$u_N \succcurlyeq v_M \leftrightarrow f(\sum_{i \in N} g(u_i)) + \sqrt[1/n]{\prod_{i \in N} I_+(u_i)} \geq f(\sum_{i \in M} g(v_i)) + \sqrt[1/m]{\prod_{i \in M} I_+(v_i)},$$

---

[9] Note that the implication of combining Avoidance of the Sadistic Conclusion and Extended Continuity seems similar but different from the so-called *mere addition* axiom. The mere addition axiom requires that the addition of one individual with zero well-being has no effect on social welfare, which easily leads to the repugnant conclusion. As I proved in Lemma 1, the combination of Avoidance of the Sadistic Conclusion and Extended Continuity implies a very similar property to the mere addition axiom, while it is different in the strict sense. The difference is that when one person with zero well-being is added to the original profile, it could be better or worse than the original profile. However, this effect vanishes after the addition of one individual with zero well-being. Once a profile has an individual with zero well-being, then any addition of individuals with zero well-being has no impact on social welfare. Hence, the egalitarian view would disagree with this property of the possibility result in Theorem 2. Also, note that the mere addition axiom implies Monotonicity for the Addition of Zero Well-being but not vice versa. Theorem 1 shows that combining the very similar but different property of the mere addition axiom and the monotonicity condition (a weaker form of the mere addition axiom) yields the repugnant conclusion.

where $f$ is a continuous and strictly monotonic function, which has an upper bound, $g$ is a continuous, strictly monotonic, and concave function with $g(0) = 0$, and $I_+$ is a function such that $I_+(u_i) = u_i$ if $u_i \geq 0$ and $I_+(u_i) = 0$ otherwise.

By definition, this SWO obviously satisfies the axioms of Avoidance of the Sadistic Conclusion, Avoidance of the Repugnant Conclusion, Anonymity, Pigou-Dalton Transfer, Strong Pareto, and Extended Continuity. ∎

The value function of the social welfare ordering in Theorem 2 is the sum of a simple transformation of generalized utilitarianism with an upper bound and a variant of a geometric mean. Under profiles with at least one zero or negative well-being, the specific form of the geometric mean is set to take on a value of zero.[10] Since both the function $f$ and the geometric mean are continuous, this social welfare ordering satisfies Extended Continuity. If one wants to concern about relative inequality, $f$ should rely on rank-weighted utilitarianism.[11]

As seen from Theorem 1, this social welfare ordering cannot satisfy Monotonicity for the Addition of Zero Well-being. Indeed, consider a profile such that it takes values close to the upper bound of the function $f$ since everyone has sufficiently high positive well-being. Adding one individual with zero well-being to this profile would result in a zero value for the geometric mean and decrease social welfare than before. As a result, no matter how high everyone else has a level of well-being, it will not be able to offset the decrease in the geometric mean, since it will only asymptotically equal the value of the upper bound of *f*.

---

[10] This means that an addition of one individual with non-positive well-being to any positive profile may greatly depreciate social welfare. This may seem a discontinuous situation. However, in fact, the addition of one individual yields a new situation of profile comparisons different from the original population size. Note that extended continuity simply requires that both the upper and lower contour sets (consisting of profiles with the same population size) are closed.

[11] Note that this social welfare ordering can avoid several other undesirable conclusions besides the repugnant and sadistic conclusions. For example, this SWO can avoid the reverse repugnant problem (the problem that a profile of one individual with sufficiently high well-being is deemed socially better than a profile of many individuals with slightly lower but sufficiently high well-being).

The next theorem shows that dropping Extended Continuity makes a possibility result, in which an SWO satisfies Monotonicity for the Addition of Zero Well-being.

*Theorem 3. There exists an SWO that satisfies Avoidance of the Sadistic Conclusion, Avoidance of the Repugnant Conclusion, Anonymity, Pigou-Dalton Transfer, Strong Pareto, and Monotonicity for the Addition of Zero Well-being.*

Proof. Consider the following SWO. For all profiles $u_N, v_M \in U$,

$u_N \succcurlyeq v_M \leftrightarrow [\sum_{i \in N: u_i \leq 0} g(u_i) > \sum_{i \in M: v_i \leq 0} g(v_i)]$ or $[\sum_{i \in N: u_i \leq 0} g(u_i) = \sum_{i \in M: v_i \leq 0} g(v_i)$ & $f(n_{++})/n_{++} \sum_{i \in N: v_i > 0} g(u_i) \geq f(m_{++})/m_{++} \sum_{i \in M: v_i > 0} g(v_i)]$,

where *g* is a continuous, strictly monotonic, and concave function with $g(0) = 0$, *f* is a strictly monotonic, and concave function, and $n_{++} = |\{i \in N | u_i > 0\}|$ & $m_{++} = |\{i \in M | v_i > 0\}|$.

This SWO is a simple lexicographic composition of a negative part of generalized utilitarianism and a positive part of number-dampened generalized utilitarianism (Ng 1986). By definition, this SWO obviously satisfies the axioms of Avoidance of the Sadistic Conclusion, Avoidance of the Repugnant Conclusion, Anonymity, Pigou-Dalton Transfer, Strong Pareto, and Monotonicity for the Addition of Zero Well-being. ∎

As Theorems 2 and 3 show, there are social evaluation rules that satisfy a part of standard axioms (Anonymity, Pigou-Doulton Transfer, and Strong Pareto) and avoid the two undesirable conclusions. The consequence of impossibility theorem arises only from the imposition of both Extended Continuity and Monotonicity for the Addition of Zero Well-being.

However, these possibility results seem not reasonable but unacceptable. For example, consider a "*galactic utopia*" with trillions of happy people, which has a very high value according to the SWO in Theorem 2. This value can be easily ruined by adding just one single person with zero well-being. Also, in Theorem 3, adding one single person with sufficiently small negative well-being makes social welfare worse than before, no matter how large the happy people improve well-being. These scenarios refer to a kind of fragility or discontinuity, which is undesirable and normatively implausible.

Moreover, as Greaves (2017) pointed out, there is another problem in population ethics, that is known to the *weak repugnant conclusion*.[12] The weak repugnant conclusion says that there exists a well-being level (e.g. a well-being level just above a critical level) such that for any profile of people having (arbitrarily high) positive well-being, a profile of enormous populations with this well-being level is socially better than the positive well-being profile.[13] Let us impose the following axiom to avoid this weak version of the repugnant conclusion.

---

[12] Blackorby and Donaldson (1984) proposed critical-level generalized utilitarianism to avoid the repugnant conclusion. But this SWO leads to the weak repugnant conclusion.

[13] If a critical level is very large (i.e. a very happy life), then there is no reason why a sufficiently large population where all individuals enjoy their very happy lives should not be better than some other population. Hence, note that the weak repugnant conclusion is only normatively compelling when the critical level is small. However, whenever the critical level is very large, another significant problem occurs as I will show later. For simplification of the notation, any constraint on the range of critical levels is not imposed in the definitions of some axioms.

***Avoidance of the Weak Repugnant Conclusion.*** $\nexists c > 0, \ \forall u_N \in U_{++}, \ \exists m \in \mathbb{N}, \ m * c \succ u_N.$[14]

By definition, Avoidance of the Weak Repugnant Conclusion implies Avoidance of the Repugnant Conclusion, but not vice versa. For example, consider a critical-level generalized utilitarian SWO with its critical level of 1. Then, this SWO satisfies Avoidance of the Repugnant Conclusion but not Avoidance of the Weak Repugnant Conclusion because any positive well-being profile is socially worse than a profile of some populations with 1 or more well-being, while any profile with sufficiently small well-being $\varepsilon$ is worse than a positive profile with 1 or more well-being.

Next, let us strengthen Avoidance of the Sadistic Conclusion. These conditions require that adding a better profile to a population is socially at least as good as adding a worse profile to it. In other words, these conditions interpret the fact that adding a bad profile to a population is better than adding a good profile to it as the *sadistic* conclusion. Let us now consider the axioms corresponding to the following two cases.

***Avoidance of the Weak Sadistic Conclusion.*** $\forall u_N, v_M, s_L \in U$ with $v_M \succ s_L$, $(u_N, v_M) \succcurlyeq (u_N, s_L)$.

***Strong Avoidance of the Weak Sadistic Conclusion.*** $\forall u_N, v_M, s_L \in U$ with $v_M \succcurlyeq s_L$, $(u_N, v_M) \succcurlyeq (u_N,$

---

[14] Note that this definition is equivalent to the following condition.
$\nexists c > 0, \forall \varepsilon > 0, \ \forall u_N \in U_{++}, \ \exists m \in \mathbb{N}, \ m * (c + \varepsilon) \succ u_N.$
In general, a critical level $c$ of a profile $u_N$ is given by $u_N \sim (u_N, c)$. This level could be variable. For example, the simple average utilitarian SWO has no fixed critical level.

$s_L$). Moreover, if $v_M > s_L$, then $(u_N, v_M) > (u_N, s_L)$.

By definition, Strong Avoidance of the Weak Sadistic Conclusion implies Avoidance of the Weak Sadistic Conclusion. Moreover, note that Avoidance of the Weak Sadistic Conclusion implies Avoidance of the Sadistic Conclusion if any positive well-being profile is socially better than any negative well-being profile.[15] As the following lemma shows, Strong Avoidance of the Weak Sadistic Conclusion can be shown to be equivalent to the *utility independence* axiom in population ethics.

Lemma 2. *An SWO $\succcurlyeq$ satisfies Strong Avoidance of the Weak Sadistic Conclusion if and only if* $\forall$ $u_N, v_M, s_L \in U, v_M \succcurlyeq s_L \leftrightarrow (u_N, v_M) \succcurlyeq (u_N, s_L)$.

Proof. A sufficiency part of the proof is trivial and I omit it. I will only prove the necessary part of this lemma.

Suppose to the contrary, that is, $(u_N, v_M) \succcurlyeq (u_N, s_L)$ but $s_L > v_M$ for some $u_N, v_M, s_L \in U$. By Strong Avoidance of the Weak Sadistic Conclusion, $s_L > v_M$ implies $(u_N, s_L) > (u_N, v_M)$ for all $u_N \in U$. However, this contradicts $(u_N, v_M) \succcurlyeq (u_N, s_L)$. ∎

The condition in Lemma 2 is called the *utility independence* axiom (Blackorby et al. 2005). It is obvious that if an SWO satisfies this independence axiom, then it satisfies Strong Avoidance of the Weak Sadistic Conclusion. Hence, from Lemma 2, they are equivalent. By using Lemma 2, I will show that attempting to avoid the weak sadistic conclusion leads to the weak repugnant conclusion. To prove this impossibility result, let us introduce the following axiom that for some profile, adding one person with some well-being to it is indifferent from the original profile.

---

[15] For example, if an SWO satisfies Minimal Increasing and its critical-level is zero, this property is easily established.

**Weak Existence of Critical Levels.** $\exists u_N \in U, \exists c \in \mathbb{R}, u_N \sim (u_N, c)$.

The above condition is a requirement for minimal comparability of variable populations. However, imposing this axiom would make the weaker versions of the repugnant and sadistic conclusions incompatible.

*Theorem 4. There exists no SWO that satisfies Strong Avoidance of the Weak Sadistic Conclusion, Avoidance of the Weak Repugnant Conclusion, Minimal Equity, Minimal Increasing, and Weak Existence of Critical Levels.*

Proof. By Lemma 2, if an SWO satisfies Strong Avoidance of the Weak Sadistic Conclusion, then it satisfies the utility independence axiom.

Then, by Weak Existence of Critical Levels, $\exists u_N \in U, \exists c \in \mathbb{R}, u_N \sim (u_N, c)$.

The utility independence axiom implies that, for all $v_M$ in $U$, $(u_N, v_M) \sim (u_N, c, v_M)$.

Because of $u_N \backsim u_N$, the utility independence axiom also implies that $v_M \backsim (c, v_M)$.

Hence, for any $v_M$ in $U$, for any natural number $k$, $v_M \backsim (k * c, v_M)$.

Minimal Equity implies that $(k + m) * \frac{1}{k+m}[kc + \sum_{i \in M} v_i] \succcurlyeq (k * c, v_M)$.

Since $\lim_{k \to \infty} \frac{1}{k+m}[kc + \sum_{i \in M} v_i] = c$, for all $\varepsilon > 0$, there exists $k' \in \mathbb{N}$, $c + \varepsilon > \frac{1}{k'+m}[k'c + \sum_{i \in M} v_i]$.

By Minimal Increasing, $(k' + m) * (c + \varepsilon) \succ (k' + m) * \frac{1}{k'+m}[k'c + \sum_{i \in M} v_i]$.

Hence, for all $v_M$ in $U$, $(k' + m) * (c + \varepsilon) \succ (k' + m) * \frac{1}{k'+m}[k'c + \sum_{i \in M} v_i] \succcurlyeq (k' * c, v_M) \backsim v_M$.

Transitivity implies $(k' + m) * (c + \varepsilon) \succ v_M$.

This holds true if $\varepsilon = 1$, which contradicts Avoidance of the Weak Repugnant Conclusion. ∎

Theorem 4 is simply proved by the fact that strong avoidance of the weak version of the sadistic conclusion has the same effect as the utility independence axiom. When the independence axiom is combined with the existence of a critical-level for additional individual well-being, the critical-level for just one profile is extended to whole critical-levels for any profile. As a result, it becomes impossible to avoid the weak repugnant conclusion. Someone may think that the weak repugnant conclusion is not *repugnant* very much because a large population where every well-being is slightly above a critical-level could be socially desirable as long as the critical-level is sufficiently high. But note that the a sufficiently high critical level must be a double-edged sword. In general, the following result holds true irrespective of the critical level.

Proposition 1. *If an SWO that satisfies Strong Avoidance of the Weak Sadistic Conclusion, Minimal Increasing, and Weak Existence of Critical Levels, then $\exists c \in \mathbb{R}, \forall \varepsilon > 0, \forall n, m \in \mathbb{N}, n * c \succ m * (c - \varepsilon)$.*

Proof. By proof of Theorem 4, if an SWO satisfies Strong Avoidance of the Weak Sadistic Conclusion and Weak Existence of Critical Levels, then there exists $c \in \mathbb{R}$ such that, for any $v_M$ in $U$, for any natural number $k$, $v_M \sim (k * c, v_M)$.

Hence, $\forall n, m \in \mathbb{N}, n * c \sim m * c$.

By Minimal Increasing, $\forall \varepsilon > 0, m * c \succ m * (c - \varepsilon)$.

By Transitivity, $n * c \succ m * (c - \varepsilon)$. ∎

Proposition 1 says that any profile with well-being slightly below the critical level must be worse than any profile with the same well-being of the critical level. This is so terrible if the critical level is sufficiently high, *n* is sufficiently small, and *m* is enormous. I shall call this problem *the reversed repugnant conclusion with a critical level*. Hence, given a fixed critical level, if the level is sufficiently low, then the weak repugnant conclusion matters. On the contrary, if the level is sufficiently high, then the reversed repugnant conclusion with a critical level matters.

Now, let us consider theoretical relationships of the weak versions of undesirable conclusions and standard axioms again. The impossibility result of Theorem 4 can be avoided by requiring Avoidance of the Weak Sadistic Conclusion instead of Strong Avoidance of the Weak Sadistic Conclusion.

*Theorem 5. There exists an SWO that satisfies Avoidance of the Weak Sadistic Conclusion, Avoidance of the Weak Repugnant Conclusion, Anonymity, Pigou-Dalton Transfer, Strong Pareto, and Weak Existence of Critical Levels.*

Proof. Consider an extended version of the leximin SWO. For all profiles $u_N, v_M \in U$,

$$u_N \succcurlyeq v_M \longleftrightarrow \begin{cases} u_N \succcurlyeq_L (v_M, (n-m) * \max_{i \in M} v_i), & \text{if } n > m, \\ (u_N, (m-n) * \max_{i \in N} u_i) \succcurlyeq_L v_M, & \text{otherwise,} \end{cases}$$

where $\succcurlyeq_L$ is the simple leximin ordering. By definition, this SWO satisfies the axioms of Avoidance of the Weak Sadistic Conclusion, Avoidance of the Weak Repugnant Conclusion, Anonymity, Pigou-Dalton Transfer, Strong Pareto, and Weak Existence of Critical Levels. ∎

As is clear from properties of lexicographic orderings, this SWO in Theorem 5 is not continuous. However, the following result shows that it is impossible to add Extended Continuity to a system of these axioms. Indeed, as the following lemma shows, combining Extended Continuity and Strong Pareto with Avoidance of the Weak Sadistic Conclusion is equivalent to the utility independence axiom.

Lemma 3. *If an SWO $\succcurlyeq$ satisfies Avoidance of the Weak Sadistic Conclusion, Strong Pareto, and Extended Continuity, then it satisfies Strong Avoidance of the Weak Sadistic Conclusion.*

Proof. By Lemma 2, I will prove that an SWO satisfies the independence axiom whenever it satisfies the above conditions. Suppose to the contrary, that is, $v_M \succ s_L$ but $(u_N, s_L) \succcurlyeq (u_N, v_M)$ for some $u_N, v_M, s_L \in U$.

By Avoidance of the Weak Sadistic Conclusion, $v_M \succ s_L$ implies $(u_N, v_M) \succcurlyeq (u_N, s_L)$. Hence, $(u_N, s_L) \sim (u_N, v_M)$ must be true.

By $v_M \succ s_L$ and Extended Continuity, for some $v'_M$ with $v'_M < v_M$, $v'_M \succ s_L$.

By Strong Pareto, it holds $(u_N, v_M) \succ (u_N, v'_M)$.

By Avoidance of the Weak Sadistic Conclusion, $v'_M \succ s_L$ implies that $(u_N, v'_M) \succcurlyeq (u_N, s_L)$.

Therefore, it holds $(u_N, s_L) \sim (u_N, v_M) \succ (u_N, v'_M) \succcurlyeq (u_N, s_L)$, that contradicts Transitivity of SWOs. ∎

From Lemma 3 and Theorem 4, the following impossibility result is immediately proved. Therefore, any continuous, efficient, and equitable SWO must lead to either the weak repugnant conclusion or the weak sadistic conclusion.

*Theorem 6. There exists no SWO that satisfies Avoidance of the Weak Sadistic Conclusion, Avoidance of the Weak Repugnant Conclusion, Minimal Equity, Strong Pareto, Extended Continuity, and Weak Existence of Critical Levels.*

Note that Strong Pareto is required in Theorem 6. Strong Pareto plays an important role of strict monotonicity, which makes Lemma 3 valid. Indeed, if only Weak Pareto (the requirement that social welfare is improved whenever everyone is strictly better-off) is required as an efficiency axiom, then the maximin criterion satisfies all the other axioms in Theorem 6. However, the maximin criterion does not appear to be acceptable, since it does not satisfy Strong Pareto and only barely satisfies various weak requirements. Hence, examining the maximin criterion is not a fruitful task and is therefore omitted from this paper. What is important is that under the assumption of Weak Existence of Critical Levels, there is no continuous, efficient, and equitable social welfare ordering that would avoid both weak versions of the repugnant and sadistic conclusions.

Note also that a possibility result can be obtained by dropping Weak Existence of Critical Levels. However, this possibility result is not at all interesting, which is indeed repugnant in the truest sense of the word. In fact, the following SWO satisfies all the standard axioms (Anonymity, Pigou-Dalton Transfer, Strong Pareto, and Extended Continuity), and avoids two undesirable conclusions. However, it should be clear to all that this is absolutely an unacceptable SWO.

*Theorem 7. There exists an SWO that satisfies Strong Avoidance of the Weak Sadistic Conclusion, Avoidance of the Weak Repugnant Conclusion, Anonymity, Pigou-Dalton Transfer, Strong Pareto, and Extended Continuity.*

Proof. Consider the following SWO. $\exists c \in \mathbb{R}$, for all profiles $u_N, v_M \in U$,

$u_N \succcurlyeq v_M \leftrightarrow [m > n]$ or $[n = m \,\&\, \sum_{i \in N}[g(u_i) - g(c)] \geq \sum_{i \in M}[g(v_i) - g(c)]\,]$.

where $g$ is a continuous, strictly monotonic, and concave function with $g(0) = 0$.

By definition, this SWO satisfies Anonymity, Pigou-Dalton Transfer, Strong Pareto, and Strong Avoidance of the Weak Sadistic Conclusion. Also, this SWO satisfies Avoidance of the Weak Repugnant Conclusion since $(\max_{i \in N} u_i + 1) \succ u_N$ for all $u_N$.

Since this SWO is generalized utilitarian whenever population sizes are the same between profiles, it is continuous for all the same-population profiles. In addition, because its upper or lower contour sets must be universal sets of the different-population profiles in comparisons of profiles with different population sizes, this SWO satisfies Extended Continuity.[16]  ∎

Theorems 1-7 show that two undesirable conclusions can be superficially avoided without abandoning the standard axioms. In an essential sense, however, it does not appear that an acceptable social welfare ordering (*Parfit's Theory X*) has been obtained. It is not surprising that in comparisons of different population sizes, there are critical levels for well-being levels of additional populations. Obviously, it is desirable to add new happy individuals to the world who have a sufficient quality of life compared to the existing standard of living. Therefore, the axiom regarding the critical levels of additional individuals should not be abandoned. Standard axioms such as continuity, efficiency, and equity would be essential to a class of acceptable social welfare orderings.

---

[16] As Blackorby et al. (2001) shows, it is known that if an SWO satisfies Weak Pareto and Extended Continuity, it can be represented by a reduced functional form of two variables, its population size and representative value of a profile. In fact, the social welfare ordering of Theorem 7 can obtain the following reduced functional form.

$u_N \succcurlyeq v_M \leftrightarrow \text{Arctan} \sum_{i \in N}[g(u_i) - g(c)] - (n-1)\pi \geq \text{Arctan} \sum_{i \in M}[g(v_i) - g(c)] - (m-1)\pi$.

In light of the results I have presented, it seems that there are two classes of acceptable social welfare orderings in population ethics. One class respects the independence axiom (i.e. Avoidance of the Weak Sadistic Conclusion), which allows situations to lead to the weak repugnant conclusion (e.g. Blackorby and Donaldson (1984)'s critical-level generalized utilitarian SWOs). The other class accepts violations of the independence axiom to avoid the weak repugnant conclusion (e.g. Pivato (2020)'s rank additive SWOs, and Sakamoto and Mori (2021)'s stepwise rank-dependent SWOs). Which class is more appropriate as an acceptable social welfare ordering would require further exploration.

# 4. Conclusion

In this study, I showed that it is possible to superficially avoid both of the two undesirable conclusions in population ethics. However, I also showed that avoiding both the repugnant and sadistic conclusions does not lead to reasonable and acceptable solutions, but to the impossibility theorems by slightly strengthening both of them. In this sense, this study confirm once again that it is important to balance the axioms in population ethics by choosing, rejecting, and weakening them, since it is not possible to satisfy all desirable axioms in social choice problems with variable populations. The remaining issues in this study are as follows.

First, it is important for any acceptable SWO to be represented by reduced functional forms. Blackorby et al. (2001) showed that an SWO that satisfies Extended Continuity and Weak Pareto simply focuses on just two variables: a population size of a profile and its representative well-being value. Since both Extended Continuity and the Pareto principle are reasonable, all acceptable social welfare orderings in population ethics should be within this class.

Second, theoretical relationships to other undesirable conclusions should be examined. For example, the reversed repugnant conclusion (Blackorby et al. 1998) is the problem that a profile of many individuals with almost zero negative well-being is socially worse than a profile of individuals with sufficiently large negative well-being. If one wishes to avoid this conclusion, the SWO in Theorem 2 could be modified as follows.

$$\forall u_N, v_M \in U,\ u_N \succcurlyeq v_M \leftrightarrow \text{Arctan} \sum_{i \in N} g(u_i) + \sqrt[\frac{1}{n}]{\prod_{i \in N} I_+(u_i)} - \sqrt[1/n]{\prod_{i \in N} I_-(u_i)} \geq \text{Arctan} \sum_{i \in M} g(v_i) + \sqrt[\frac{1}{m}]{\prod_{i \in M} I_+(v_i)} - \sqrt[1/m]{\prod_{i \in M} I_-(v_i)},$$

where $g$ is a continuous, strictly monotonic, and concave function with $g(0) = 0$, $I_+$ is a function such that $I_+(u_i) = u_i$ if $u_i \geq 0$ and $I_+(u_i) = 0$ otherwise, and $I_-$ is a function such that $I_-(u_i) =$

$u_i$ if $u_i < 0$ and $I_-(u_i) = 0$ otherwise

The value function of the above social welfare ordering is the sum of a simple transformation of generalized utilitarianism with its upper and lower bounds and variants of the geometric mean. Clearly from the definition, both the original and reversed versions of the repugnant conclusions are avoidable. Several other undesirable conclusions are known in general, but I will not discuss them further.

Third, while this study has shown that the strong version of avoidance of the sadistic conclusion and the independence axiom are equivalent, different versions could be considered. For example, the following conditions could be considered.

*Addition of Indifferent Profiles.* $\forall u_N, v_M, s_L \in U$ with $v_M \sim s_L$, $(u_N, v_M) \sim (u_N, s_L)$.

*Positive Responsiveness of Adding Profiles.* $\forall u_N, v_M, s_L \in U$ with $v_M \succ s_L$, $(u_N, v_M) \succ (u_N, s_L)$.

Combining the two conditions above implies the independence axiom, and thus leads to the same impossibility result. If a society requires only one of these conditions independently, some possibility and impossibility results are obtained, but I omit them because it is not very fruitful.

Fourth, it seems important to consider the relevancy of the *unrestricted repugnant conclusion*, a modified version of the repugnant conclusion proposed by Spears and Budolfson (2021). They consider the result to be *repugnant*, that for some population, adding an enormous population where each member has almost zero well-being to the population is socially better than adding a population where each member has sufficiently high well-being to it. Formally, their unrestricted repugnant conclusion is defined as follows.

***The Unrestricted Repugnant Conclusion.*** $\forall u_N \in U_{++}$, $\forall \varepsilon > 0$, $\exists k \in \mathbb{N}$, $\exists v_M \in U \cup \{\emptyset\}$, $(k * \varepsilon, v_M) \succ (u_N, v_M)$.

However, while the original repugnant conclusion is contrary to intuition, the unrestricted repugnant conclusion does not necessarily seem to be contrary to intuition. Also, their impossibility theorem depends critically on the conditions they call the *sign axioms*, which are not satisfied by many of the classes of acceptable SWOs. In this sense, ethical implications of the unrestricted repugnant conclusion and the similar undesirable conclusion should be reexamined.[17]

Finally, the basic and central problem of population ethics appears to be reducible to the question of the extent to which a society should accept a change in well-being levels of the original population for an additional population. The core of this contention may be the separability axiom. Whether a situation in which an additional population has too low a level of well-being relative to the well-being level of the existing population should be considered good or not would require further consideration in population ethics. On the other hand, as the repugnant conclusion illustrates, in order to add an enormous population where every member has almost zero well-being, the level of well-being of the existing population should not be reduced to any extent. Since this balancing problem does not appear to be solvable by any axiomatic analysis, it will need to be tested by various economic

---

[17] Their impossibility result focuses on the *very repugnant conclusion* which is closely similar to the unrestricted repugnant conclusion. Formally, the very repugnant conclusion is defined as follows.

$\forall u_N \in U_{++}$, $\forall s_L \in U_{--}$, $\forall \varepsilon > 0$, $\exists k \in \mathbb{N}$, $\exists v_M \in U \cup \{\emptyset\}$, $(k * \varepsilon, s_L, v_M) \succ (u_N, v_M)$.

and philosophical experiments.